\documentclass[12pt,a4]{article}
\usepackage{graphicx}
\usepackage{latexsym,amssymb}
\usepackage{amsmath,amsfonts,amsthm}
\usepackage{epstopdf}
\usepackage{booktabs}

\title{A parallel algorithm for the enumeration of benzenoid hydrocarbons}
\author{Iwan Jensen \\ Department of Mathematics and Statistics \\ 
The University of Melbourne, Vic. 3010, Australia}
 
\begin{document}

\maketitle

\begin{abstract}
We present an improved parallel algorithm for the enumeration of fixed benzenoids
$B_h$ containing $h$ hexagonal cells. We can thus extend the enumeration of $B_h$ 
from the previous best $h=35$ up to $h=50$.  Analysis of the associated generating
function confirms to a very high degree of certainty that $B_h \sim A \kappa^h /h$ and
we estimate that the growth constant $\kappa = 5.161930154(8)$ and the amplitude 
$A=0.2808499(1)$. 
\end{abstract}


\section{Introduction \label{sec:intro}}

A benzenoid or planar polyhex is a special type of hydrocarbon
molecule. Its hexagonal system is obtained by deleting
all carbon-hydrogen bonds, leaving clusters of hexagons
joined at an edge (a carbon-carbon bond). They thus appear
as  clusters of identical hexagons in the plane. 
The interior of the clusters are filled with hexagons so there
are no internal holes. These structures have appeared independently
in the chemical and mathematical literature.
In the mathematics literature they are discussed as self-avoiding
polygons on the hexagonal lattice \cite{Enting89} and a distinction is 
made between fixed and free embeddings. Fixed polygons
are considered distinct up to a translation while free polygons are considered
equivalent under translations, rotations and reflections. 
Polygons are typically enumerated according to their perimeter or area.
In the chemistry literature the number of free polygons \cite{GutmanCyvin} has
been universally considered. The number of benzenoids 
or planar polyhexes is equal to the number of free hexagonal self-avoiding polygons
enumerated by area.

The enumeration of the number $b_h$ of benzenoids of $h$ cells 
remains an important topic in computational and theoretical chemistry.
The monograph by Gutman and Cyvin \cite{GutmanCyvin} provides a 
comprehensive review of all aspects. Until a few years ago progress was slow
and  incremental as calculations were based on direct counting of benzenoids.
As the number of these grows as $b_h \sim  \kappa^h$, where the
growth constant $\kappa \simeq 5.16$, it is clear that,
to obtain one further term one needs more than 5 times the
computing power. Up to 1989, the number of benzenoids up to $h = 12$ was 
known \cite{GutmanCyvin}. Ten years later this had been improved to 
$h = 21$ \cite{Caporossi98}, while more recently, the number of benzenoids up
to $h = 24$ was obtained \cite{Brinkmann01}.  
In 2002 \cite{Voge02} a major break-through was obtained using a different
type of algorithm that enabled the number of {\em fixed} benzenoids $B_h$ to 
be enumerated for $h \leq 35$ and $b_h$ was then obtained to the same size by  
using direct counting algorithms to enumerate benzenoids possessing certain 
symmetries, e.g. they may be symmetric with respect to an axis of reflection or
certain rotations.  For direct counting algorithms the CPU time  taken to
enumerate $B_h$ grows as $\kappa^h$, whereas for our algorithm time consumption
grows approximately  as $1.65^h$, since $1.65<\kappa \simeq 5.16$ we may say
that the new algorithm is exponentially faster than direct counting. 
Its drawbacks are that it is much more memory intensive 
(memory grows exponentially with $h$)
than direct counting, for which memory requirements are
negligible, as well as being much more difficult to implement.

In \cite{Voge02} it was shown that there exists a growth constant $\kappa$ 
such that
\begin{equation}\label{eq:growth}
\lim_{h\to \infty} B_h^{1/h}=\kappa
\end{equation}
and the universally accepted, but as yet unproved, conjecture
\begin{equation}\label{eq:asymp}
B_h \sim A\kappa^h h^\theta  \,\,\, {\rm as} \,\,\, h\to \infty
\end{equation}
for the asymptotic form for $B_h$ was confirmed to a high degree of certainty.
It is widely accepted that for models such as benzenoids, other self-avoiding
polygon models enumerated by area and polyominoes (or lattice animals) 
the exponent $\theta$ is given by the Lee-Yang edge singularity exponent \cite{Parisi81}
and thus $\theta=-1$ for benzenoids. Numerical analysis \cite{Voge02} 
confirmed this conjecture to a very high degree of certitude and yielded
the estimate $\kappa = 5.16193016(8)$ for the growth constant and
$A=0.2808491(1)$ for the critical amplitude.

In this paper we describe an efficient parallel version of the 
algorithm used in \cite{Voge02} and extend the count for fixed
benzenoids up to $h=50$. We do not attempt to count $b_h$ since
asymptotically $B_h = 12b_h$ so any results regarding the asymptotic
behaviour of $B_h$ and $b_h$ are essentially the same (and the
ratio of the two sequences $B_h/b_h$ converge rapidly to its asymptotic limit 
as evidenced by the fact that $12-B_{35}/b_{35}\simeq 1.355 \times 10^{-10}$).
Furthermore the direct counting algorithms for benzenoids with a symmetry
have computational complexity $\lambda^h$ where $\lambda=\kappa^{1/k}$
if enumerating benzenoids with a $k$-fold symmetry so that in the worst case 
we have $\lambda = \sqrt{\kappa} \simeq 2.27$, which is a much worse 
asymptotic growth than that achieved with the algorithm for fixed benzenoids.
Our analysis of the extended data yields the even more precise estimates
$\kappa=5.161930154(8)$ and a revised estimate for the critical amplitude
$A=0.2808499(1)$. 

\section{Computer algorithm \label{sec:algo}}

A detailed description of the original computer algorithm can be found in \cite{Voge02}.
For this work we use a slightly different algorithm and we have therefore chosen 
to describe it in some detail below before specifying how it can be turned
into an efficient parallel algorithm.

\subsection{Finite lattice algorithm \label{sec:flm}}

\begin{figure}
\begin{center}
\includegraphics[scale=0.8]{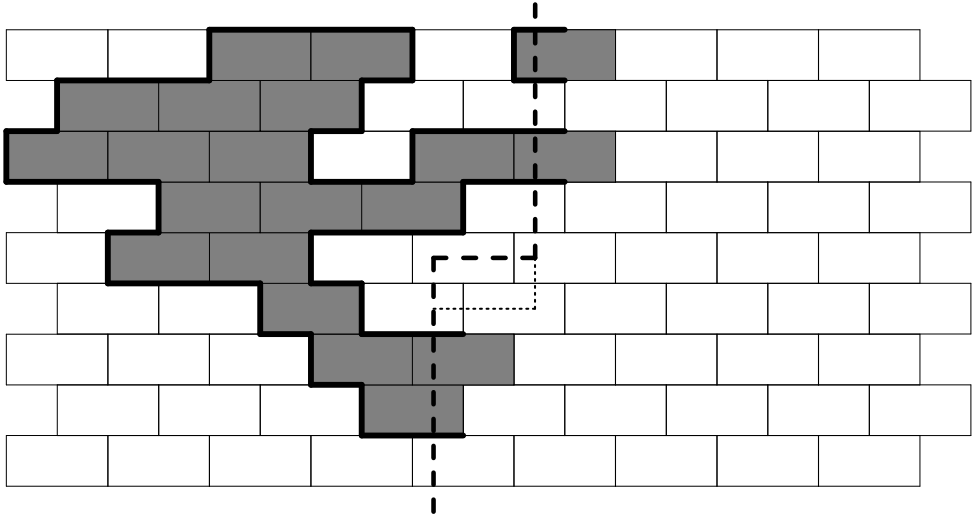}
\end{center}
\caption{\label{fig:transfer}
A snapshot of the boundary line (dashed line) during the transfer matrix 
calculation on the brick-work lattice. Benzenoids are enumerated by successive
moves of the kink in the boundary line, as exemplified by the position given 
by the dotted line, so that two vertices at a time are added to the rectangle. 
To the left of the boundary line we have drawn (shaded cells) an example of a 
partially completed benzenoid.}
\end{figure}

We count the number of fixed benzenoids using the so-called finite lattice
method pioneered by Enting \cite{Enting80}. In this method the number of
benzenoids are obtained by calculating the contributions from benzenoids
contained within finite sub-lattices. As in \cite{Enting89,Voge02} we embed
the hexagonal lattice  in the square lattice as the brick-work lattice 
(see Fig.~\ref{fig:transfer}) and our finite lattices are rectangles of width $W$ 
and length $L$. The minimum number of cells needed to span a rectangle from top to
bottom and left to right is essentially $W+\max (0,L- (W+1)/2 )$ 
(simply note that a single `line' of cells starting in the top-left corner and going
down the diagonal contains $W$ cells and extends $ (W+1)/2$ cells 
to the right). So benzenoids up to a maximal size $h_{\rm max}$ can be counted
be combining the counts from all finite $W\times L$ lattices with 
$W+\max (0,L- (W+1)/2) \leq h_{\rm max}$.

The number of benzenoids in a given rectangle is calculated using transfer-matrix 
techniques. The transfer matrix (TM) technique  involves drawing a boundary line through 
the rectangle intersecting a set of up to $W+1$ edges. Benzenoids in a given rectangle 
are enumerated by moving the boundary line so as to add two vertices (or a single cell)
at a time as shown in Fig.~\ref{fig:transfer}. In this fashion we build up the rectangle
column by column with each column built up cell by cell.  As we move the  boundary line 
it intersects partially completed benzenoids consisting of disjoint loops that must all be 
connected to form a single benzenoid. This TM algorithm is used for
rectangles where $L\geq W$. Note that the hexagonal lattice (or bricklayer lattice)
is not symmetric with respect to rotation. So for rectangles with $L<W$ we choose instead 
to let the boundary line cut across $L+1$ edges in the length wise direction and we
then move the boundary line from the bottom to the top of the rectangle. This ensures
that the number of edges cut by the boundary line is minimal  and at most
$2h_{\rm max}/3$. The TM algorithm in the two cases are essentially identical and 
differ only in `surface' effects. Below we give some further details of the TM algorithm. 

To avoid situations leading to graphs with more than a single
component we have to forbid a loop to close on itself if the boundary line 
intersects any other loops. So two loop ends can only be joined if they belong 
to different loops or all other edges are empty. To exclude loops which close
on themselves we need to label the occupied edges in such a way that
we can easily determine whether or not two loop ends belong to the same loop.
The most obvious choice would be to give each loop a unique label.
However, on two-dimensional lattices there is a more compact scheme
relying on the fact that two loops can never intertwine. Each end of a loop 
is assigned one of two labels depending on whether it is the lower end or 
the upper end of a loop. Each configuration along the boundary line can thus 
be represented by a set of edge states or a state vector $s=\{\sigma_i\}$, where

\begin{equation}\label{eq:sapstates}
\sigma_i  = \left\{ \begin{array}{rl}
 0 &\;\;\; \mbox{empty edge},  \\ 
 1 &\;\;\; \mbox{lower end of a loop}, \\
 2 &\;\;\; \mbox{upper end of a loop}. \\
\end{array} \right.
\end{equation}
\noindent
With this encoding the state along the boundary line in Fig.~\ref{fig:transfer} is
$s=\{01010002212\}$. It is easy to see that this encoding uniquely describes 
which loop-ends are connected. In order to find the upper loop-end, matching a 
given lower end, we start at the lower end and work upwards  in the configuration 
counting the number of `1's and `2's we pass (the `1' of the initial lower end is 
{\em not} included in the count). We stop when the number of `2's  exceeds the 
number of  `1's. This `2'  marks the matching upper end of the loop.

\begin{figure}
\begin{center}
\includegraphics[scale=1.0]{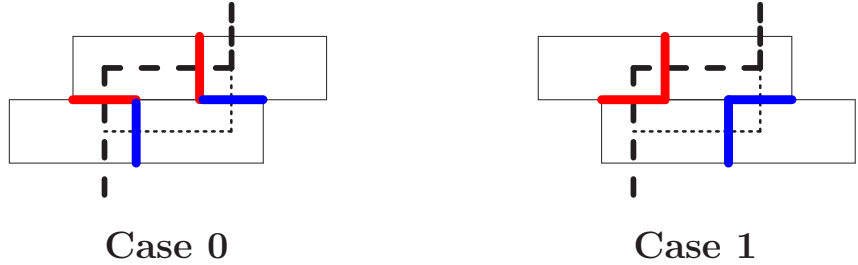}
\end{center}
\caption{\label{fig:case}
The two different update cases encountered in the move of the TM boundary line.
Red (blue) edges indicate the kink edges before (after) the move.
}
\end{figure}

When the boundary line is moved we encounter two different cases as we add
a new cell as illustrated in Fig.~\ref{fig:case}. When building up a new column we
alternate between the two cases. For each configuration of occupied or 
empty edges along the boundary, we maintain a generating function for partially
completed benzenoids. The generating function is a (truncated)
polynomial $p_s(q)$, where $s$ is the state vector specifying
the `source' configuration.  When the boundary line is moved, a given state vector
$s$ is transformed into two new state `target' vectors $t_1$ and $t_2$ and
$q^{k_1}p_s(q)$ is added to $p_{t_1}(q)$ and $q^{k_2}p_s(q)$ is added to $p_{t_2}(q)$, 
where $k_1$ and $k_2$ are 1 or 0 depending on whether the new cell is 
part of the benzenoid or not. It is quite simple to determine
whether a newly added unit cell  belongs to a benzenoid or not. Moving 
through a configuration we note that as we reach the first occupied 
edge we pass from the outside to the inside of a benzenoid, the next occupied
edge takes us to the outside again, and so on. In this fashion
all unit cells intersected by the boundary line are uniquely
assigned to the interior or exterior of a benzenoid.

\begin{figure}[ht]
\begin{center}
\includegraphics[scale=0.7]{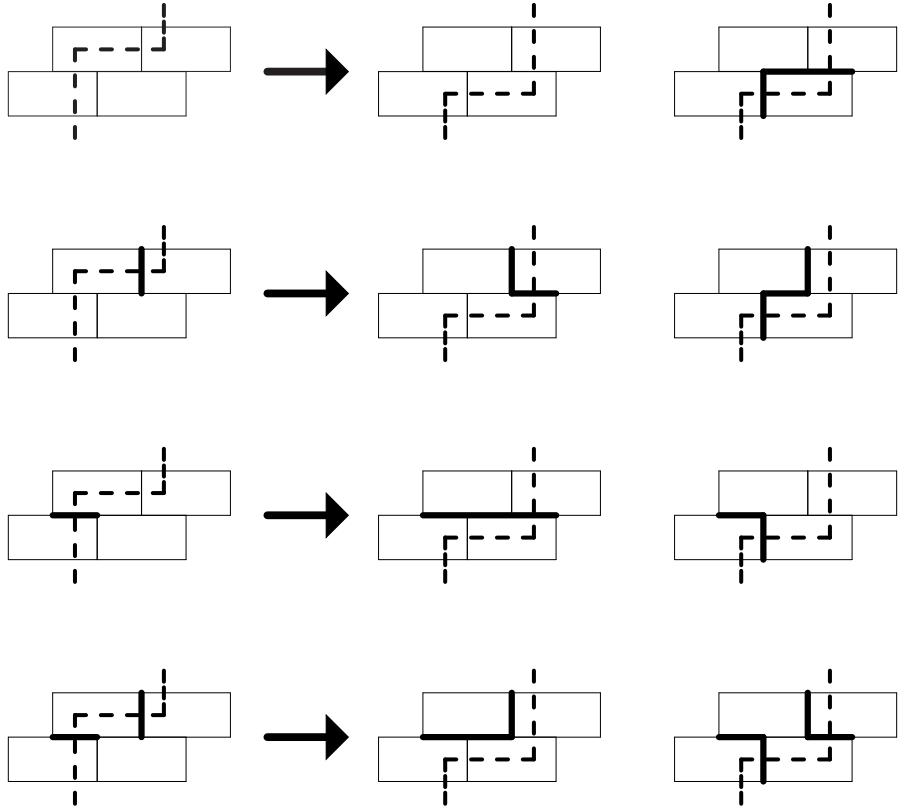}
\end{center}
\caption{\label{fig:case0}
The possible updates in Case 0 when the input state (left-most column) has 0, 1
or 2 occupied edges. The right-most columns shows the possible outputs. }
\end{figure}

\begin{figure}[ht]
\begin{center}
\includegraphics[scale=0.7]{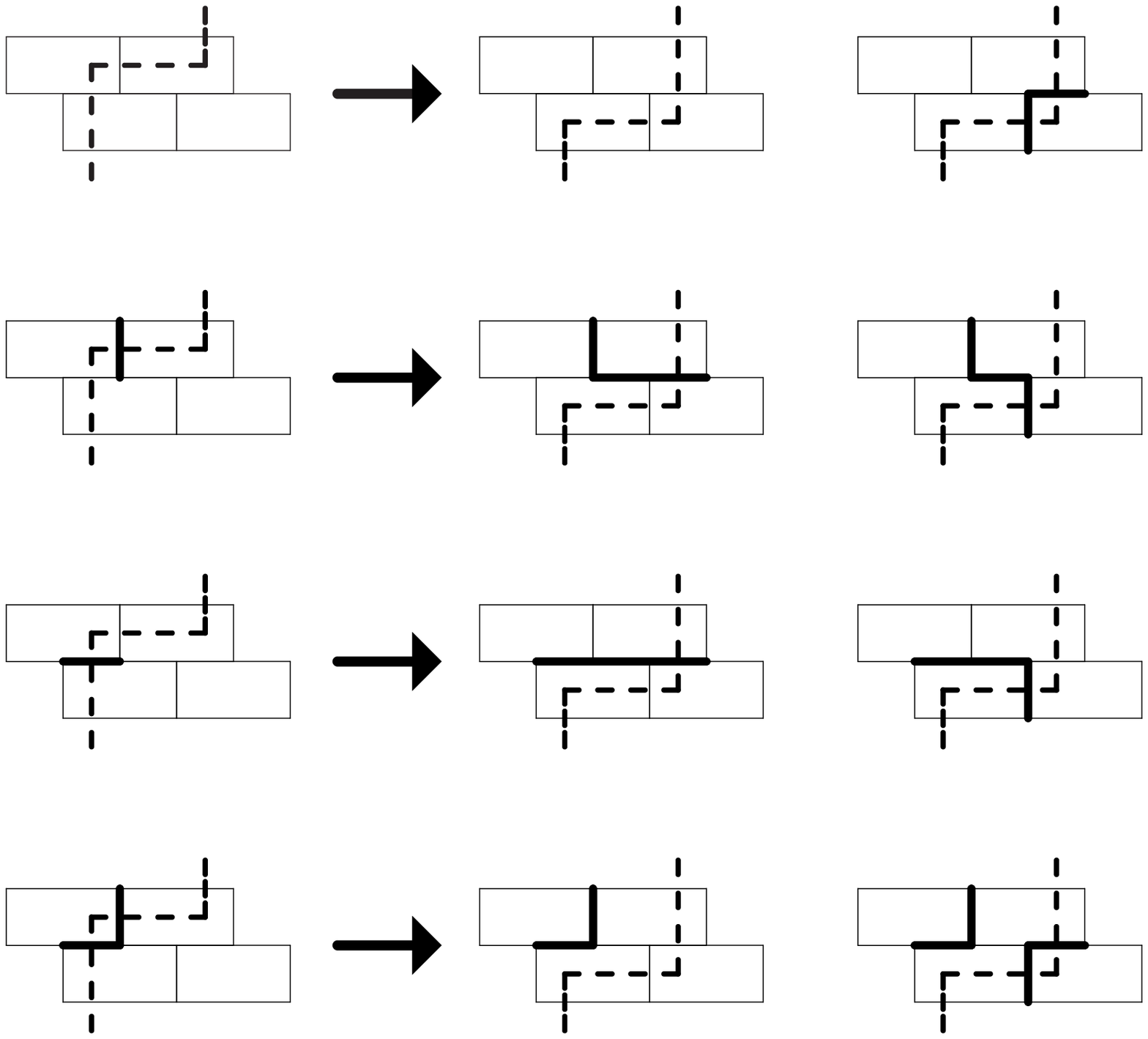}
\end{center}
\caption{\label{fig:case1}
Similar to Fig.~\ref{fig:case0} but for Case 1.
}
\end{figure}

In Fig.~\ref{fig:case0} and \ref{fig:case1} we illustrate the possible new configurations
of the edges in the kink of the boundary line as we add a new cell. The actual  update rules
will depend not only on the number of occupied kink edges in the input configuration
but on their states as well. The update rules are summarised in Table~\ref{tab:update}
and a few comments are in order. The first five rows should be self-explanatory. 
In rows six and nine over-lining of the output state means that we have connected
two lower (upper) loop-ends and we therefore have to relabel one of the matching upper
(lower) loop-ends in the target state as a lower (upper) state. The matching loop-ends
are easily located as explained below (\ref{eq:sapstates}). In row seven Acc means 
accumulate into final count for $B_h$ if valid. Here we are forming a closed loop and
this is only allowed if there are no other occupied edges in the state (otherwise we
either produce  graphs with several separate components or interior holes neither
of which are permissible benzenoids). In Case 1 row seven the second output can never occur.
Finally in row eight we connect upper and lower loop-ends from two different
loops. This is always allowed and the outputs states need no further comments.

\begin{table}
\caption{ \label{tab:update}
Update rules for Case 0 and Case 1}
\begin{center}
\begin{tabular}{lllllll}
\toprule
 \multicolumn{3}{c}{Case 0} & \phantom{xxxx}  &  \multicolumn{3}{c}{Case 1}  \\
 \midrule
Input & \multicolumn{2}{c}{Output} &  & Input & \multicolumn{2}{c}{Output}  \\
\midrule
`00'  & `00' & `12' & & '00'   & `00' & `'12' \\
`01'  & `01' & `10' & & `01'  & `01' & `10' \\
`02'  & `02' & `20' & & `02'  & `02' & `20' \\
`10'  & `01' & `10' & & `10'  & `01' & `10' \\
`20'  & `02' & `20' & & `20'  & `02` & `20' \\
`11'  & `$\overline{00}$' & `11' & & `11'  & `$\overline{00}$' & `$\overline{12}$' \\
`12'  & Acc & `12' & & `12'  & Acc & -----\\
`21'  & `00' & `21' & & `21` & `00' & `12' \\
`22'  & `$\overline{00}$' & `22' & & `22`  & `$\overline{00}$'  & `$\overline{12}$'  \\
\bottomrule
\end{tabular}
\end{center}
\end{table}

A major improvement to the basic method can be obtained by using
the approach first adopted in \cite{Jensen99}. As stated earlier we require valid 
benzenoids to span the rectangle in {\em both} directions. In other words we directly 
enumerate benzenoids of width exactly $W$ and length $L$. To implement the TM
algorithm efficiently we use several memory and time saving methods. The most 
important is what we call {\em pruning}. This procedure, details of which are given
in \cite{Jensen99}, allows us to discard most of the possible
configurations for large $W$ because they only contribute to benzenoids of 
size greater than $h_{\rm max}$. Briefly this works as follows.``
For each configuration we keep track of the current minimum number of 
cells $h_{\rm cur}$ already inserted to the left of the boundary line. We then
calculate the minimum number of additional cells $h_{\rm add}$ required to 
produce a valid benzenoid. There are three contributions, namely the number 
of cells required to close the benzenoid, the number of cells needed (if any) 
to ensure that the benzenoid touches both the lower and upper border, and 
finally the number of cells needed (if any) to extend at least $W$ cells 
in the length-wise direction (remember we are looking at rectangles
with $L \geq W$). If the sum $h_{\rm cur}+h_{\rm add} > h_{\rm max}$ we 
can discard the partial generating function for that configuration,
and of course the configuration itself, because it won't make a 
contribution to the benzenoid count up to the size we are 
trying to obtain.

Those familiar with algebraic languages will recognize that each configuration of labeled
loop-ends forms a Motzkin word \cite{Delest84}. It is known that the number of Motzkin words
of length $m$ grows like $3^m$. The maximal number of bonds intersected
by the boundary line grows as $2h_{\rm max}/3$. This implies that the
complexity of enumerating benzenoids of size $h$ grows as
$3^{2h/3}\simeq   2.08^ h$, multiplied by some polynomial in $h$. Thus the basic 
transfer-matrix approach already provides a dramatic improvement over
direct enumeration algorithms, which have complexity $5.16^h$. With the further
improvements outlined above, it is not possible to give a theoretical analysis of the 
computational complexity of the algorithm, but an empirical analysis in \cite{Voge02} suggested
that the improvements reduce the complexity to $\lambda^h$ with $\lambda \simeq  1.65$.
For this work a slight further improvement has been obtained reducing
$\lambda$ to 1.56 or so. In addition some further memory saving strategies were 
adopted. The effectiveness of these can be gauged by noting that in \cite{Voge02}
the calculation of $B_h$ up to $h=35$ required some 5Gb of memory and we can 
now achieve a similar task using only some 250Mb of memory.

The integers $B_h$ become very  large and exceed $2^{64}$ which causes overflow
when using 64-bit integers. The solution to this problem is use modular arithmetic and do 
the calculation modulo several numbers $p_i$ and then reconstruct the true $B_h$ using 
the Chinese remainder theorem \cite{Knuth97}. In our case it sufficed to do the calculations
modulo  $p_0=2^{62}$ and $p_1=2^{62}-1$. It should be noted that the computationally
expensive part of our algorithm is pruning. Compared to this the time taken to perform
the modular calculations updating the partial generating functions is insignificant.
Since the calculations were done on a shared facility CPU time was more of a premium 
than memory and we did the calculation using both $p_0$ and $p_1$ in a single run.
Total CPU time expended on the calculations was approximately 22000 CPU hours.

In Table~\ref{tab:series} we list the additional 15 terms for $B_h$ with $h \geq 36$
obtained in this work, the original 35 terms can be found in \cite{Voge02} or 
down-loaded from our web-site \cite{JensenWeb}. 
\begin{table}[bt]
   \centering
   \caption{ \label{tab:series}
    Number of fixed benzenoids $B_h$ of size $h\geq 36$.}
  \begin{tabular}{@{} ll @{}} 
      \toprule
 $h$ & $B_h$ \\
  \midrule     
36  &  352506828543839738006802 \\
37   &  1771125269041561567830953  \\
38   &  8905113919188230264955009  \\
39   &  44804571829235959198699855  \\
40   &  225570974088699920561748746  \\
41   &  1136340745302289809680018862  \\
42   &  5727773558054438208070950886  \\
43   &  28887056504374868913302241736  \\
44   &  145763914212751560334802981991  \\
45   &  735894997233174457602406978869  \\
46   &  3716988842355112053567240722854  \\
47   &  18783102592560998779533576292617  \\
48   &  94958908613774943408509332060260  \\
49   &  480273434248924455452231252618009  \\
50   &  2430068453031180290203185942420933  \\
      \bottomrule
   \end{tabular}
\end{table}

\subsection{Parallelisation \label{sec:para}}

The computational complexity of the FLM grows exponentially with
the number of terms one wishes to calculate. It is therefore 
little wonder that implementations of the algorithms have always
been geared towards using the most powerful computers available. By now
parallel computing is well established as the paradigm
for high performance computing and in particular cluster computing has
emerged as the dominant platform for large scale computing facilities.
The transfer-matrix algorithms used in the calculations of the
finite lattice contributions are eminently suited for parallel
computations. 

The most basic concerns in any efficient parallel algorithm is
to minimise the communication between processors and ensure that
each processor does roughly the same amount of work and use similar amounts
of memory. In practice one naturally has to strike some compromise
and accept a certain degree of variation across the processors.

One of the main ways of achieving a good parallel algorithm using 
data decomposition is to try to find an invariant under the
operation of the updating rules. That is we seek to find some property
about the configurations along the boundary line which
does not change in a single iteration.
The algorithm for the enumeration of benzenoids is quite complicated 
since not all possible configurations occur due to pruning
and an update at a given set of edges might change the state of 
an edge far removed, e.g., when two lower loop-ends are joined
we have to relabel one of the associated upper loop-ends as
a lower loop-end in the new configuration.
However, there still is an invariant since any edge not
directly involved in the update cannot change from being 
empty to being occupied and vice versa. That is only the edges 
at the kink of the boundary line can change their occupation status. This invariant
allows us to parallelise the algorithm in such a way
that we can do the calculation completely independently on each
processor with just two redistributions of the 
data set each time an extra column is added to the lattice. 

The main points of the algorithm are summarized below:

\begin{enumerate}
\item With the boundary line straight (having no kinks) distribute the
configurations and their generating functions across processors so that 
configurations with the same occupation pattern along the {\em lower} half 
of the boundary line are placed on the same processor. 
\item Do the TM update inserting the top-half of a new column.
This can be done {\em independently} by each processor because the 
occupation pattern in the lower half remains unchanged.
\item Upon reaching the half-way mark redistribute the data
so that configurations with the same occupation pattern along 
the {\em upper} half of the boundary line are placed on the same processor. 
\item Do the TM update inserting the bottom-half of a new column.
\item Go back to 1.
\end{enumerate}

The redistribution among processors was done as follows: 

\begin{enumerate}
\item On each processor run through the configurations to establish
the occupation pattern (in the lower or upper half of the boundary) $c$ of each 
configuration and calculate, $n(c)$, the number of configurations with a given pattern.
\item Calculate the {\em global} sum of $n(c)$.
\item Sort the global sum $n(c)$.
\item Assign each pattern to a processor $p_i$ as follows:
\begin{enumerate} 
\item Set $p_i=0$.
\item Assign the {\em most} frequent unassigned pattern $c$ to processor $p_i$. 
\item If the number of configurations assigned to $p_i$ is
less than the number of configurations assigned to processor 0 then
assign the {\em least} frequent  unassigned patterns to $p_i$ until
the desired inequality is achieved.
\item Set $p_i = (p_i+1) \mod N_p$, where $N_p$ is the number of processors.
\item Repeat from (b) until all patterns have been assigned.
\end{enumerate}
\item On each processor run through the configurations sending
each configuration to its assigned processor. 
\end{enumerate}

\begin{table}
\caption{ \label{tab:para}
Number of processors with total CPU-time and 
actual running time (in the format hh:mm) 
as well and memory use for the parallel algorithm for
enumerating benzenoids of maximal size 43 at width 22.}
\begin{center}
\begin{tabular}{rrrrrrr}
\toprule
Proc. & Total time  & Run time  & Max Conf & Min Conf & Max Term & Min Term \\
\midrule
1     &  60:13  &   60:20   &  107350066 &          &  207111142 & \\
2     &  61:53   &    30:59   &52982622 &  52435395 & 102711198 &  102666398 \\
4     &   62:28 & 15:38     &  26389619 & 26183924 & 51559593 &  51025667 \\
8     &  63:17   &     7:55  & 13289367 & 13078219 & 26179885 & 25492182 \\
16    &  69:28   &     4:22  &  6725270 &  6486246 & 13245615  & 12717598 \\
32    &  69:05   &     2:10  &  3440269 &  3274193 &  6871820 &  6347966 \\
64    &  71:33   &       1:08  &  1768626 &  1616220 & 3839775 & 3191842    \\
\bottomrule
\end{tabular}
\end{center}
\end{table}

The bulk of the calculations were performed on the facilities of the Australian
Partnership for Advanced Computing (APAC). The APAC facility is an
SGI Altix cluster with 1920 1.6 Ghz  Itanium2 processors grouped into 30 partitions
with 64 processors each. The cluster has a total peak speed over 11Tflops. 
Nodes are connected via a SGI's NUMAlink with a latency $<$ 2 us (MPI) and 
bandwidth  of 3.2 Gb/sec bidirectional. We used up to 128 processors per run
using a maximum of 230Gb of memory and 22000 CPU hours.

In Table~\ref{tab:para} we have listed the time and memory use of the
algorithm for $h_{\rm max} = 43$ at $W=22$ using from 1 to 64 processors.
The memory use of the single processor job was about 3Gb. Firstly, we
look at the issue of balancing the memory use  of the 
parallel algorithm. By design we are attempting to balance this to the
greatest extend possible since in a cluster environment  memory is often
the most crucially constrained resource. This aspect
is examined via the numbers in columns 4--7. At any given time during the 
calculation each processor handles a subset of the total number of configurations.
For each processor we monitor the maximal number of configurations and
terms retained in the generating functions. 
The  balancing can be roughly gauged by looking at the
largest (Max Conf) and smallest (Min Conf) maximum number of 
configurations handled by individual processors during the execution of 
the program. In columns 6 and 7 are listed the largest (Max Term) and
smallest (Min Term) number of terms retained in the generating
functions associated with the subset of configurations. 
As can be seen the algorithm is quite well balanced. Even with 64 processors, where each
processor uses only about 50Mb of memory, the difference between the processor
handling the maximal and minimal number of configurations is less than 10\%.
For the total number of terms retained in the generating functions the difference is 
less than 20\%.  So our aim of balancing memory use has clearly been met.

The next issue is that of balancing the CPU time used by the algorithm. As can be seen 
the algorithm scales reasonable well from 1 to 64 processors since the total combined 
CPU time (column 2, format is hours:minutes) used by all processors increase only by 
about 10\%. Likewise the run time (column 3, format is hours:minutes:seconds)
of the program is approximately halved when the number of processors is doubled. 
This is not quite as good a scaling as achieved for some previous algorithms
\cite{Jensen03,IJ03b} where the total CPU time stayed constant. The main reason
for the discrepancy is that the time consuming part of our algorithm is the pruning.
For ``simpler" problems on the square lattice  it turned out that the
time consumption was fairly constant irrespective of the occupation pattern.
Pruning benzenoid configurations is more complicated\footnote{We won't give details here
but just note that  on the square lattice the three contributions to $h_{\rm add}$ essentially
de-couple and can be determined more or less independently. This is no longer
the case on the hexagonal lattice vastly complicating the pruning.}. In our previous
work \cite{Jensen03,IJ03b} the CPU time used in communication tasks never
exceeded 10\% of the total. However, for benzenoids a simple timing of the
various routines show that as much as 30\% of the time was used in communication
task. We believe most of the additional time use is due to `latency'. That is the task
of redistributing the data among processors must complete before further processing
can be done. The redistribution is thus blocking. If certain subsets of configurations
sitting on processor $p_j$ take long to process they can thus lead to imbalances 
where other processors must wait for the completion of the calculation on processor
$p_j$. Unfortunately it is not possible to determine a priori if a certain set of 
configurations with a particular occupation pattern are `slow'. However, it does 
suggest that there is some room for improvement to the redistribution, perhaps by 
including additional information (say which borders have been touched or the total 
number of occupied edges) so as to further sub-divide the set of configurations thus 
making it easier to balance the workload. Another option would be to monitor the 
time used to process each configuration and use this as part of the information 
used in the redistribution. However, this should not come at the cost of unbalanced 
memory use. These possibilities remain to be explored in future work.

\section{Numerical analysis \label{sec:ana}}

From the coefficients 
$B_h$  we have the first 50 terms in the respective generating function,
\begin{equation}\label{eq:Gf}
G(q) = \sum_{h} B_h q^h \sim A(q) (1- \kappa q) ^{-1-\theta}
\end{equation}
where the functional form of the generating follows from (\ref{eq:asymp}) with
the radius of convergence of the generating function given by $q_c=1/\kappa$ . 
In order to obtain the singularity structure of the
generating function we used the numerical method of
differential approximants \cite{GuttmannDA}. Very briefly, in this method we
approximate the generating function by the solution to a
linear, inhomogeneous, ordinary differential equation (ODE)
with polynomial coefficients The singular behaviour
of such ODEs is a well known classical mathematics problem 
(see e.g. \cite{Ince27}) and the singular points and
exponents are easily calculated. Even if the function {\em globally} is not a solution
of a such a linear ODE (as is the case for SAP) one hopes that {\em locally} in the
vicinity of the (physical) critical points the generating function can still be well
approximated by a solution to a linear ODE.

A $K^{th}$-order differential approximant (DA) to a function $F(x)$  is formed by matching 
the coefficients in the polynomials $Q_i(x)$ and $P(x)$ of degree $N_i$ and $L$, respectively,
so that (one) of the formal solutions to the inhomogeneous differential equation
\begin{equation} \label{eq:ana_DA}
\sum_{i=0}^K Q_{i}(x)(x\frac{{\rm d}}{{\rm d}x})^i \tilde{F}(x) = P(x)
\end{equation}
agrees with the first $M=L+\sum_i (N_i+1)$ series coefficients of $F$.
We normalise the DA by setting $Q_K(0)=1$ thus leaving us with $M$ rather
than $M+1$ unknown coefficients.
The singularities of $F(x)$ are thus approximated by zeroes $x_i$ of $Q_K(x)$ and the 
associated critical exponent $\lambda_i$ is estimated from the associated indicial 
equation \cite{Ince27}. 

One can increase the degree of the polynomials
and the order of the underlying differential equation until
there are no more known coefficients. For each specific choice of order and degrees
one must then solve a set of linear equation for the coefficients in the
polynomials in the approximating ODE. A
substantial number of such differential approximants are
constructed, and a statistical procedure is used to estimate
the critical point and critical exponent. All calculations were
carried out using floating point arithmetic with quadruple precision
(our algorithm has been tested extensively using Maple with 
precision of 100 digits and this revealed that rounding errors are
not an issue).

\begin{table}[t]
   \centering
   \caption{ \label{tab:DA}
    Estimates for the critical point $q_c=1/\kappa$ and critical exponent $-1-\theta$
     as obtained from 2nd and 3rd order differential approximants  with $L$ being
     the degree of the inhomogeneous polynomial.}
  \begin{tabular}{@{} rllll @{}} 
      \toprule
  & \multicolumn{2}{c}{2nd order approximants}   & \multicolumn{2}{c}{3rd order approximants}  \\
  \midrule
 $L$ &$q_c=1/\kappa$ & $-1-\theta$ &  $q_c=1/\kappa$ & $-1-\theta$ \\
  \midrule     
 0  & 0.19372598474(16)& -0.00000136(87)& 0.19372598440(23)& -0.00000055(37) \\
 2  & 0.19372598448(24)& -0.00000077(43)& 0.193725984286(90)& -0.00000036(16) \\
 4  & 0.19372598440(11)& -0.00000056(42)& 0.19372598436(22)& -0.00000051(39) \\
 6  &  0.19372598443(27)& -0.00000068(51)& 0.19372598416(16)& -0.00000009(41) \\
 8  & 0.19372598441(32)& -0.00000052(93)& 0.193725984182(83)& -0.00000013(21) \\
10 & 0.19372598444(19)& -0.00000069(38)& 0.193725984205(94)& -0.00000020(23) \\
      \bottomrule
   \end{tabular}
\end{table}

In Table~\ref{tab:DA} we have listed estimates for the critical point $q_c=1/\kappa$ and
critical exponent $-1-\theta$ obtained from a differential approximant analysis~\cite{GuttmannDA}.
The estimates were obtained by averaging over many individual approximants using
a procedure (see \cite{Jensen06}  for details) which automatically discard any spurious
outlying approximants. Each approximant used at least 42 terms of the series and the degree
of the inhomogeneous polynomial was varied from $L=0$ to 10.  Taken together the estimates  
are consistent with the conjectured exact value $\theta=-1$ for the critical  exponent, while for 
the critical point we obtain $q_c=0.1937259843(3)$ or for the growth constant 
$\kappa=5.161930154(8)$.

\begin{figure}
\begin{center}
\includegraphics[scale=0.85]{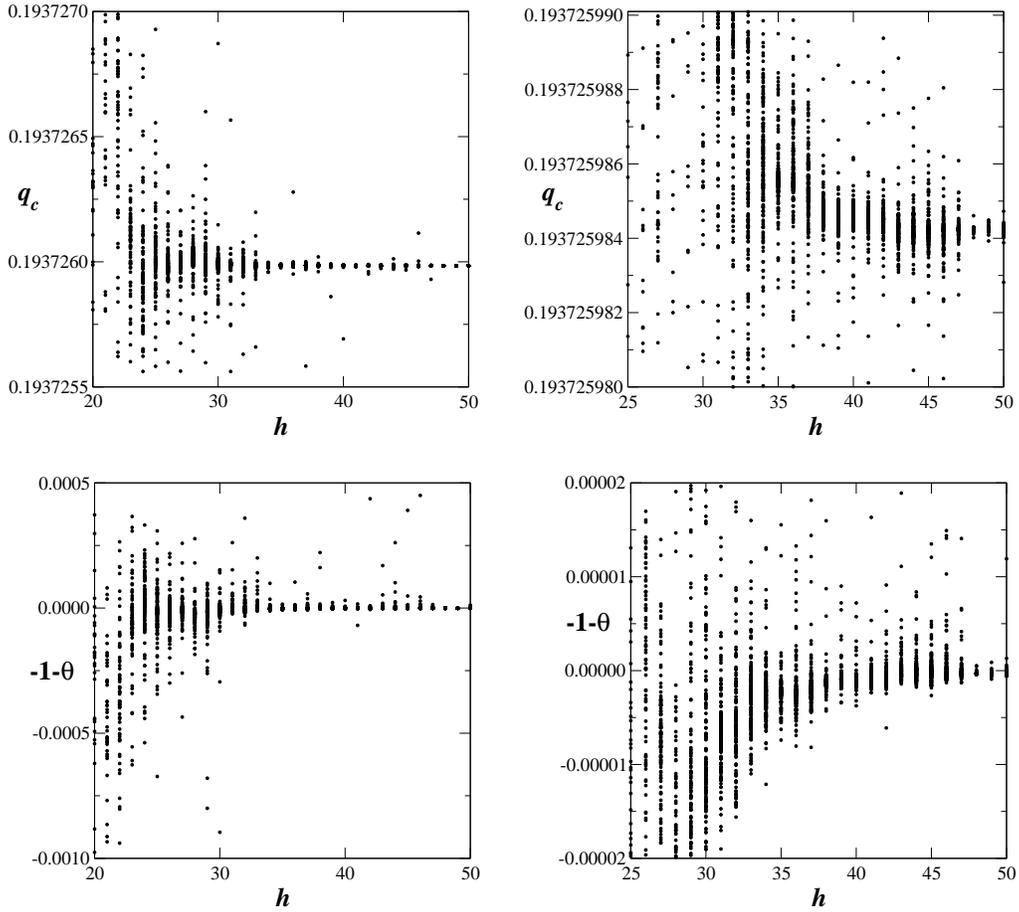}
\end{center}
\caption{\label{fig:CrpExp}
Estimates for the critical point $q_c$ (top panels) and critical exponent $-1-\theta$ (bottom panels)
vs the maximal size $h$ (or number of terms) used in the differential approximant analysis.
Each dot represents a data point obtained from a 3rd order approximant with $L=0, 2,\ldots,10$.
The left panels show a view of most approximants while the right panels are a more detailed
view at the data for high values of $h$.}
\end{figure}

While the estimates listed in Table~\ref{tab:DA}  are very accurate one issue which always
arises in a differential approximant analysis is the possibility of systematic bias. In particular 
it is possible that the estimates have not yet converged to their true asymptotic values.
In order to address this possibility we plot in Fig.~\ref{fig:CrpExp} individual estimates for 
the critical point $q_c$  and critical exponent $-1-\theta$ as a function of the maximal size or 
number of terms $h$ used to form the differential approximant. From this figure it is clear
that the estimates do settle down to very well defined values.  There is no sign of any systematic 
drift in the estimates for $h>40$ or so. In particular the conclusion
that $\theta=-1$ exactly appear to be completely safe. Likewise the estimates for $q_c$
settle down to a value in full agreement with the estimate  $q_c=0.1937259843(3)$ from
above.

\begin{figure}
\begin{center}
\includegraphics[scale=0.9]{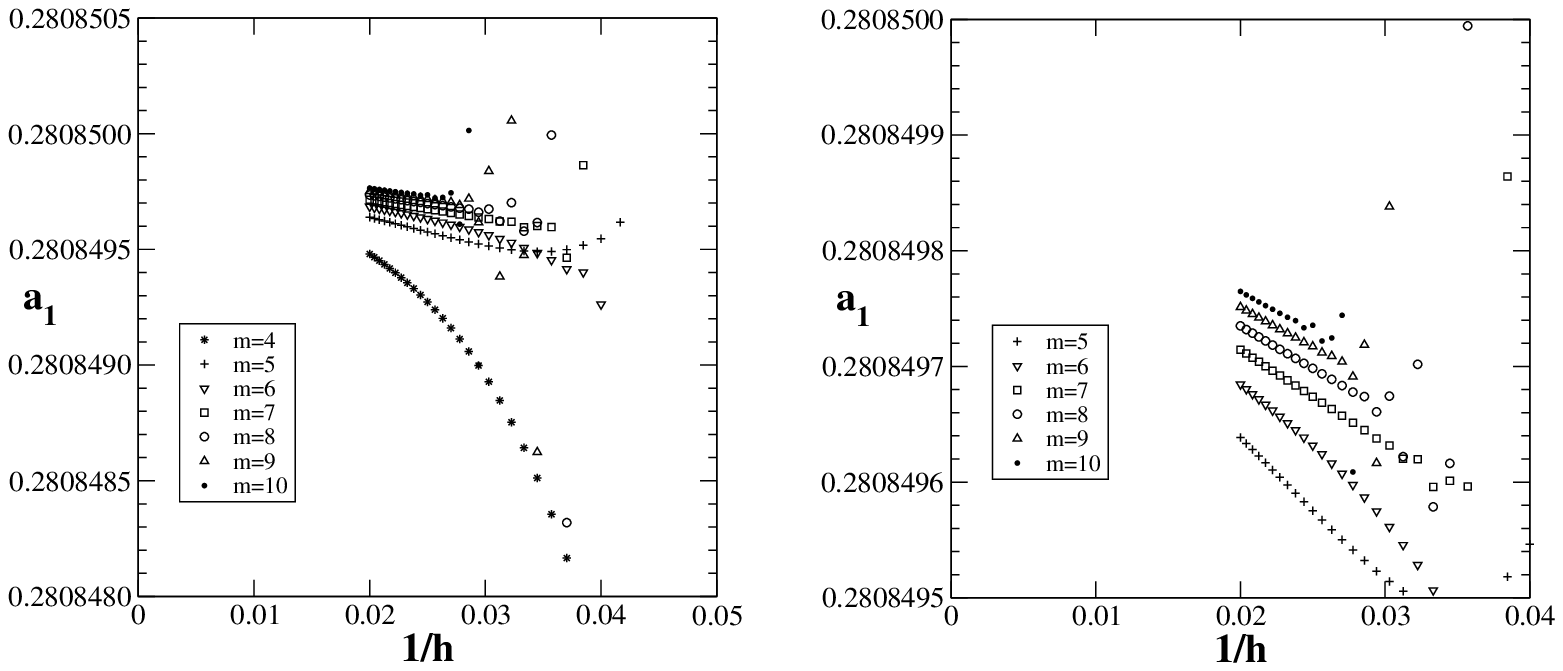}
\end{center}
\caption{\label{fig:Ampl}
Estimates  for the leading amplitude $a_1$ vs $1/h$ where $h$ is the maximal size
used in the fit to the asymptotic form (\ref{eq:Bh}) for the coefficients $B_h$. 
The plot in the right panel is a more detailed view of the data in the left panel.}
\end{figure}

Now that the exact value of $\theta$ has been confirmed and an accurate estimate for 
 $\kappa$ obtained we turn our attention to the ``fine structure'' of the asymptotic form of the
coefficients. In particular we are interested in obtaining accurate estimates for the leading 
critical amplitude  $A$. Our method of analysis consists in fitting the coefficients to an 
assumed asymptotic form. The asymptotic form (\ref{eq:asymp}) for the coefficients 
$B_h$ only explicitly gives the leading contribution. In general one would expect corrections
 to scaling given by a set of correction-to-scaling exponents. We have argued
elsewhere \cite{Jensen00a} and found in the previous study \cite{Voge02} that
there is no sign of non-analytic correction-to-scaling exponents.  The upshot
of this is that $B_h$ follows the asymptotic form
\begin{equation}\label{eq:Bh}
B_h = \kappa^h \left [ a_1/h+a_2/h^2+a_3/h^3+\cdots + {\rm O}(\exp(-h))\right ].
\end{equation}
We then obtain estimates for $a_1=A$ by fitting $B_h$ to this form. 
That is we truncate~(\ref{eq:Bh}) after $m$ terms, take a sub-sequence of coefficients
$\{B_h,B_{h-1},\ldots,B_{h-m+1}\}$, plug into the formula above and solve the 
resulting $m$ linear equations to obtain estimates for the amplitudes. It is then 
advantageous to plot estimates for the leading amplitude $a_1$ against $1/h$ for 
several values of $m$. The results are plotted in the left panel of figure~\ref{fig:Ampl}.
We clearly have very well behaved estimates. In the right panel we take a more
detailed look at the data and from this plot we estimate that $a_1=0.2808499(1)$.
In a similar manner we estimate that $a_2=-0.14518(2)$. The following amplitudes
$a_k, \,\, k\geq 3$ show quite a lot of curvature (and some even appear to diverge).
This would indicate that the asymptotic form (\ref{eq:Bh}) is in fact not quite correct
and thus that our previous conclusion that there is no sign of non-analytic
correction-to-scaling exponents may well be incorrect (at least for this problem).
We tried to include other ad hoc terms (such as half-integer exponents) but none of these
improved the convergence and essentially had no effect of the estimate for $a_1$.
The conclusion to our experimentation is that we are certain that the first two terms
of   (\ref{eq:Bh})  are correct, but beyond this more terms with non-integer exponents
are likely to occur though we do not as yet have a clear idea of the possible values
of these non-analytic correction-to-scaling exponents.

\section*{Acknowledgments}

The calculations presented in this paper would not have been possible
without a generous grant of computer time on the server cluster of the
Australian Partnership for Advanced Computing (APAC). We also used
the computational resources of the Victorian Partnership for Advanced 
Computing (VPAC). We gratefully acknowledge financial support from 
the Australian Research Council.

\end{document}